\documentclass[a4paper,12pt]{revtex4}
\usepackage{amssymb}
\usepackage{amsmath}
\usepackage{graphics}
\usepackage{color}
\usepackage[T1]{fontenc}
\usepackage{times,mathptm}
\usepackage[cp1250]{inputenc}
\usepackage{epsfig,epsf}
\newcommand{\be}{\begin{equation}}
\newcommand{\ee}{\end{equation}}

\newcommand\beq{\begin{eqnarray}}
\newcommand\eeq{\end{eqnarray}}
\begin{document}

\title{Phase diagram of the non-uniform chiral condensate in different regularization schemes at T=0}

\author{Tomasz L. Partyka and Mariusz Sadzikowski}

\affiliation{Smoluchowski Institute of Physics, Jagellonian
University, Reymonta 4, 30-059 Krak\'ow, Poland}

\begin{abstract}
We show that the qualitative picture of the phase diagram which
includes the non-uniform chiral phase and 2SC superconducting phase
is independent of the considered regularization schemes. We also
demonstrate that the quantitative results agree with each other
reasonably for the set of so called "relativistic" regularization
schemes. On the other hand the "non-relativistic" momentum cut-off
is clearly differ from the others.
\end{abstract}


\maketitle

\section{Introduction}

The phase diagram of strongly interacting matter is in the center of
scientific interest for many decades. During the last decade
there was a great progress in our understanding of this subject which
was based on the simple observation that Cooper instability and asymptotic
freedom jointly lead to the colour superconductivity phenomenon at
asymptotically high densities \cite{shuryak_alford_son}. However, at moderate
densities our knowledge is still limited to the model dependent calculations.
A variety of phases are possible which is a direct consequence of the
richness of strong (and indirectly weak) interactions between elementary
constituents of matter. In particular, the high symmetry groups involved
in the description allow the existence of many different types of phases. Of special
interest are non-uniform phases. Among those, the LOFF phases
of superconductivity \cite{loff_alford}, Overhauser effect \cite{over}
or non-uniform chiral condensates \cite{dautry_bron_sad_jap} are very good examples.

In this paper, we ask the technical but important question of the
regularization dependence of the phase diagram of strongly
interacting matter which contains the non-uniform chiral condensate.
This drawback is inferred by the fact that the NJL model describing strong
interactions is an effective non-renormalizable approximation. Then different regularization schemes lead, in a
sense, to different models. It is important to check if qualitative
results are independent of the regularization scheme, and what is their dependence
at the quantitative level.

We consider the phase
diagram at finite density which includes chiral uniform and non-uniform
phases, superconducting 2SC phase and plasma of the free quarks \cite{sad}.
Herein we checked the phase diagram within the Nambu - Jona-Lasinio model
against four various schemes, namely 3- and 4-dim cut-off,
Schwinger and Pauli-Villars regularization. Similar analysis was performed in the case
of single non-uniform chiral phase in \cite{bron_kutsch}.
As the main result we confirm that the non-uniform phase exists in all considered regularization schemes.

We expect that the most important differences emerge at zero temperature thus
we consider only this situation. At higher temperature, the major role would
be played by the finite temperature corrections. However, inclusion of those
is outside the scope of our paper.

\section{Model}

The starting point is based on the Nambu - Jona-Lasinio model with two flavours \cite{sad1}
\be
H=\int_x\left\{\bar{\psi}(i\gamma^\nu\partial_\nu +\mu\gamma_0)\psi
+G\left[ (\bar{\psi}\psi )^2+(\bar{\psi}i\gamma_5\vec{\tau}\psi )^2\right]
+G^\prime (\bar{\psi}_ci\gamma_5\tau_2\lambda^A\psi ) (\bar{\psi}i\gamma_5\tau_2\lambda^A\psi_c )\right\}
\ee
where $\psi $ is the quark field, $\psi_c=C\bar{\psi}^T$ is the conjugate field
 and $\mu $ is the quark chemical potential. The color, flavor and spinor indices are suppressed. The vector
$\vec{\tau}$ is the isospin vector of Pauli matrices and $\lambda^A$, $A=2,5,7$ are three color
antisymmetric $SU(3)$ group generators. The integration $\int_x=\int_0^\beta d\tau\int d^3x $,
where $\beta $ is the inverse temperature and derivative operator $\partial_\nu = (i\partial_\tau ,\vec{\nabla})$.
Two coupling constants $G, G^\prime $ describe interactions which are responsible for the creation of
quark-antiquark and quark-quark condensates respectively. Both couplings are treated as independent.
There is also an additional parameter $\Lambda$ which defines the energy scale below which
the effective theory applies. It is introduced through the regularization procedure.

We are working in the mean field approximation within the ansatz \cite{sad1}
\begin{eqnarray}
\label{ansatz}
\langle\bar{\psi}\psi\rangle = -\frac{M}{2G}\cos\vec{q}\cdot\vec{x},\;\;\;
\langle\bar{\psi}i\gamma_5\tau^a\psi\rangle = -\frac{M}{2G}\delta_{a3}\sin\vec{q}\cdot\vec{x},\;\;\;
\langle\psi\tau_2\lambda^AC\gamma_5\psi\rangle = \frac{\Delta}{2G^\prime}\delta_{A2}
\end{eqnarray}
which describes three possible phases: the chiral uniform phase ($\vec{q}=0, M\neq 0, \Delta =0$),
the non-uniform chiral phase ($\vec{q}\neq 0, M\neq 0, \Delta =0$) and the superconducting
phase ($\vec{q} = 0, M = 0, \Delta\neq 0$). All mentioned phases can coexist with each other.
Using standard methods one can calculate the thermodynamic potential \cite{sad, sad1}
\beq
\label{omega0}
&&\Omega_0 = \frac{M^2}{4G} +\frac{|\Delta |^2}{4G^\prime}+2\sum_{s=\pm}\int_{E_s\leq\mu}\frac{d^3k}{(2\pi)^3} (E_s-\mu)
-2\sum_{s=\pm}\int\frac{d^3k}{(2\pi)^3} \left(E_s+\sum_{i=\pm}E^\Delta_{i,s}\right),\nonumber\\
&&E^\Delta_{\pm,s}=\sqrt{(\mu\pm E_s)^2+|\Delta|^2},\;\;
E_\pm=\sqrt{\vec{k}^2+M^2+\frac{\vec{q}^{\,2}}{4}\pm \sqrt{(\vec{q}\cdot\vec{k})^2+M^2\vec{q}^{\,2}}}
\eeq
where the limit of zero temperature was already performed. The last integral is
divergent. Before we introduce different regularization schemes let us convert equation (\ref{omega0}) into another form
which much better suits our purposes and better underlines the physics of the problem.
To reach our goal we translate equation (\ref{omega0}) into another form
\beq
\label{omega}
\Omega &=& \frac{M^2}{4G} +\frac{|\Delta |^2}{4G^\prime}
-2\sum_{s=\pm}\int\frac{d^3k}{(2\pi)^3} \left(\sum_{i=\pm}E^\Delta_{i,s}-2E_s\right)
+4\sum_{i=\pm}\int\frac{d^3k}{(2\pi)^3} \left(E^\Delta_{i,0}-E^{\Delta=0}_{i,0}\right)\nonumber\\
&+& 2\sum_{s=\pm}\int_{E_s\leq\mu}\frac{d^3k}{(2\pi)^3} (E_s-\mu)
-4\sum_{i=\pm}\int\frac{d^3k}{(2\pi)^3} \left(E^\Delta_{i,0}-E^{\Delta=0}_{i,0}\right)
-6\sum_{s=\pm}\int\frac{d^3k}{(2\pi)^3}E_s, \nonumber\\
E^\Delta_{\pm,0}&=&\sqrt{(\mu\pm E_0)^2+|\Delta|^2},\;\;E_0=\sqrt{\vec{k}^2+M^2} .
\eeq
The first three integrals give finite contributions and only the last two are divergent.
Let us note that in the absence of superconducting state the next to the last term vanishes
and the only divergent contribution follows from the infinite Dirac sea integral.
Additionally the last two integrals are separately dependent, the first one on $\Delta$ and the other
one on wave vector $\vec{q}$. This separation is very convenient for the regularization procedure.

\section{Regularization schemes and parameters}

In the first step we expand the last term of equation (\ref{omega}) in powers of the wave vector $\vec{q}$.
It is well known that the parameter at the second order is related to the pion decay constant
\cite{dautry_bron_sad_jap}
\be
\label{div}
-6\sum_{s=\pm}\int\frac{d^3k}{(2\pi)^3}E_s \approx - 12\int\frac{d^3k}{(2\pi)^3} E_{0}
+\frac{M^2f^{2}_\pi \vec{q}^{\,2}}{2M^{2}_0} .
\ee
where $M_0$ is constituent quark mass at zero density.
The formula for the pion decay constant depends on the regularization and is known
from the earlier literature (e.g. \cite{klev}). This formula together with
expressions for the chiral condensate fixes the value of $G$ and $\Lambda$
parameters (Table 1).
More details are discussed in the Appendix A. The coupling
constant $G^\prime$ cannot be related to any known physical quantity.
In the vast literature of color supercoductivity, its value is emplaced
somewhere between $G/2\lesssim G^\prime\lesssim 3G/2$.
\begin{center}
\begin{table}[h]
\begin{tabular}{ccccc}
          & 3D & 4D & S & PV \\
\hline\hline
$\Lambda\;\;\;$    & 0.635 & 1.015 & 1.086 & 1.12 \\
$G\Lambda^2\;\;\;$ & 2.2 & 3.93  & 3.78  & 4.47 \\
$M_0\;\;\;$        & 0.33 & 0.238 & 0.2   & 0.22 \\
\hline\hline
\end{tabular}
\caption{Numerical values of the regularization parameter $\Lambda$ in GeV
and dimensionless quantity $G\Lambda^2$ for different regularization schemes. In the last row,
the value of the quark constituent mass at zero density $M_0$ is given in GeV.}
\end{table}
\end{center}
Taking into account equation (\ref{div}) one can extract
the divergent part of thermodynamic potential (\ref{omega}) in the form
\be
\label{omega_div}
\Omega_{div}=-4\sum_{i=\pm}\int\frac{d^3k}{(2\pi)^3} \left(E^\Delta_{i,0}-E^{\Delta=0}_{i,0}\right)
-12\int\frac{d^3k}{(2\pi)^3}E_0 .
\ee
We consider four types of different regularization schemes \cite{klev}:
\begin{itemize}
\item \textbf{3-dim cut-off (3D)} $\Lambda$ restricts the value of three dimensional momentum.
The regularized potential takes the form
\be
\label{omega_3D}
\Omega^{3D}_{div}=-4\sum_{i=\pm}\int^\Lambda\frac{d^3k}{(2\pi)^3} \left(E^\Delta_{i,0}-E^{\Delta=0}_{i,0}\right)
-12\int^\Lambda\frac{d^3k}{(2\pi)^3}E_0 .
\ee
This regularization was frequently used in previous papers e.g. \cite{sad}.
\item \textbf{4-dim cut-off (4D)} restricts the value of four-momentum vector in Euclidean space. Using the formulae
$$
\int\frac{d^3k}{(2\pi)^3}(A-B) = \int\frac{d^4k_E}{(2\pi)^4}\ln\frac{k_0^2+A^2}{k_0^2+B^2}
$$
we introduce the cut-off parameter to the thermodynamic potential through the equation
\be
\label{omega_4D}
\Omega^{4D}_{div}=-8\int^\Lambda\frac{d^4k_E}{(2\pi)^4}
\ln\frac{k_0^2+\frac{1}{4}(\sum_{i=\pm}E^\Delta_{i,0})^2}{k_0^2+\frac{1}{4}(\sum_{i=\pm}E^{\Delta=0}_{i,0})^2}
-12\int^\Lambda\frac{d^4k_E}{(2\pi)^4}\ln (k_0^2+E_0^2) ,
\ee
where $d^4k_E=dk_0d^3k$.
\item \textbf{Schwinger regularization (S)} is based on the formula
$$
\ln\frac{A}{B} = \int_0^\infty\frac{d\tau}{\tau}[\exp (\tau (B+i\epsilon))-\exp (\tau (A+i\epsilon))].
$$
This leads us to the regularized expression for the potential
\beq
\label{omega_S}
\Omega^{S}_{div}&=&-8\int\frac{d^4k_E}{(2\pi)^4}\int_{1/\Lambda^2}^\infty\frac{d\tau}{\tau}
\left\{\exp \left[\tau \left(-k_0^2-\frac{1}{4}(\sum_{i=\pm}E^\Delta_{i,0})^2\right)\right]\right.\nonumber\\
&-&\left.\exp \left[\tau \left(-k_0^2-\frac{1}{4}(\sum_{i=\pm}E^{\Delta=0}_{i,0})^2\right)\right]\right\}\nonumber\\
&-&12\int\frac{d^4k_E}{(2\pi)^4}\int_{1/\Lambda^2}^\infty\frac{d\tau}{\tau}
\left\{\exp \left[\tau \left(-k_0^2-E_0^2\right)\right]\right\}.
\eeq
This regularization was considered in \cite{nakano} where only the single non-uniform
chiral phase was taken into account.
\item \textbf{Pauli-Villars regularization (PV)} introduces an arbitrary number of
coupling constants $C_a$ and mass regulators $M_a$ combine in such a way that
divergent potential (\ref{omega_div}) becomes finite. In the first step
we regularized potential by the 3-dim cut-off and then expand the result
around the large value of the parameter $\Lambda$. The values of coupling constants $C_a, M_a$
are already set by the conditions which follows from the calculation of
the chiral condensate $\langle\bar{\psi}\psi\rangle$ and pion decay constant (the appendix).
Then final expression for the thermodynamic potential in this scheme reads
\beq
\label{omega_PV}
\Omega^{PV}_{div} &=& \frac{3}{8\pi^2}\sum_{\alpha=1}^3C_\alpha M^4_\alpha\ln\frac{M^2}{M_\alpha^2}+
\frac{1}{\pi^2}\left(\frac{|\Delta|^4}{4}-|\Delta|^2\mu^2\right)
\sum_{\alpha=1}^3C_\alpha\ln\frac{M^2+|\Delta|^2}{M_\alpha^2+|\Delta|^2}\nonumber\\
&+&\frac{|\Delta|^2}{2\pi^2}\sum_{\alpha=1}^3C_\alpha M_\alpha^2\ln\frac{M^2+|\Delta|^2}{M_\alpha^2+|\Delta|^2}
+\frac{1}{4\pi^2}\sum_{\alpha=1}^3C_\alpha M_\alpha^4
\left(\ln\frac{M^2+|\Delta|^2}{M_\alpha^2+|\Delta|^2}-\ln\frac{M^2}{M_\alpha^2}\right) ,
\eeq
\end{itemize}
where $C_a, M_a$ are given by equations (\ref{PV_param}) in the appendix.

\section{Results}

We minimize the thermodynamic potential
\beq
\label{omega_reg}
\Omega &=& \frac{M^2}{4G} +\frac{|\Delta |^2}{4G^\prime}+\frac{M^2F^{2}_\pi \vec{q}^{\,2}}{2M^{2}_0}
-2\sum_{s=\pm}\int\frac{d^3k}{(2\pi)^3} \left(\sum_{i=\pm}E^\Delta_{i,s}-2E_s\right)
+4\sum_{i=\pm}\int\frac{d^3k}{(2\pi)^3} \left(E^\Delta_{i,0}-E^{\Delta=0}_{i,0}\right)\nonumber\\
&+& 2\sum_{s=\pm}\int_{E_s\leq\mu}\frac{d^3k}{(2\pi)^3} (E_s-\mu) + \Omega^{scheme}_{div}
\eeq
with respect to mass $M$, wave vector $\left|\vec{q}\right|$ and gap parameter $\Delta$
as a function of chemical potential. The last term of (\ref{omega_reg}) depends on
the scheme as was described in the previous section.
Appropriate formule for $\Omega^{scheme}_{div}$ are given by equations
(\ref{omega_3D}, \ref{omega_4D}, \ref{omega_S}, \ref{omega_PV}).

As already discussed $G, \Lambda$ coupling constants are determined by the values
of the pion decay constant and the size of the chiral condensate at zero density
(Table I). The remaining constant $G^\prime$ is essentially unknown. For the presentation of
the result we assume $G^\prime = 0.75 G$ which follows from the Fierz transformation \cite{buballa}.
The choice of another ratio $G^\prime /G$ does not influence our analysis of the result dependence
on regularization schemes. The values of constituent mass, wave vector and gap parameter as a function
of chemical potential for different regularization schemes are given in figures 1 - 3.
As can be seen in all regularization schemes, there is the same
pattern of the phase transitions. From uniform to non-uniform chiral phase
and then to superconducting phase, all transitions are first order and existence
of non-uniform phase is then independent of the considered regularization schemes.
However, the strengths of the transition depend on the regularization scheme. This is particularly visible in Fig. 3.

\begin{figure}[t]
\begin{center}
\includegraphics[width=.8\textwidth]{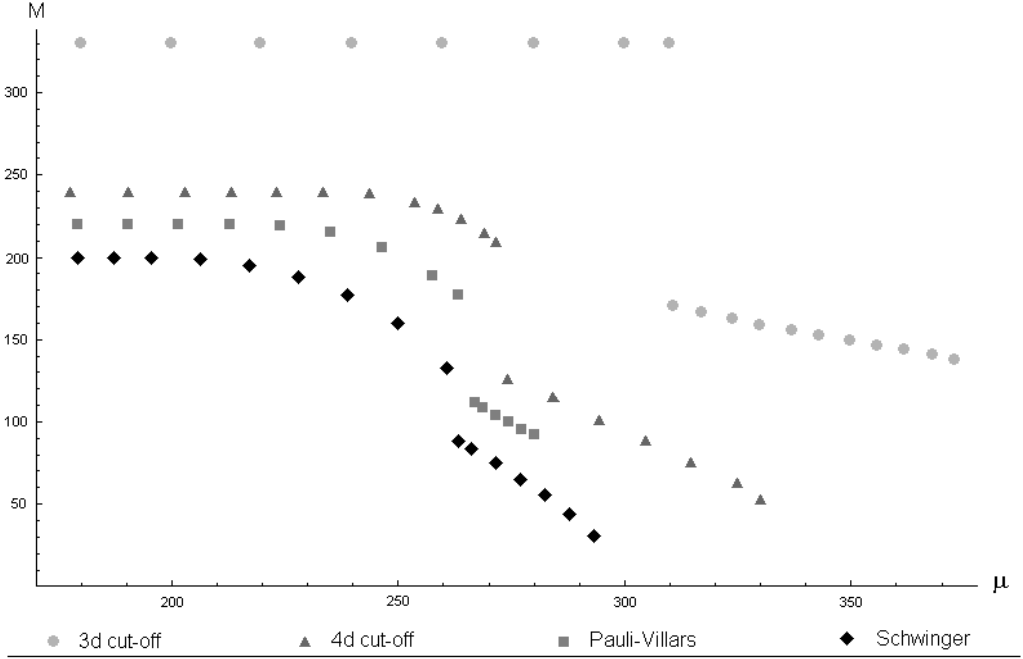}
\end{center}
\caption{The constituent mass $M$ as a function of chemical potential $\mu$
in different regularization schemes in MeV units.}
\label{number}
\end{figure}

\begin{figure}[t]
\begin{center}
\includegraphics[width=.8\textwidth]{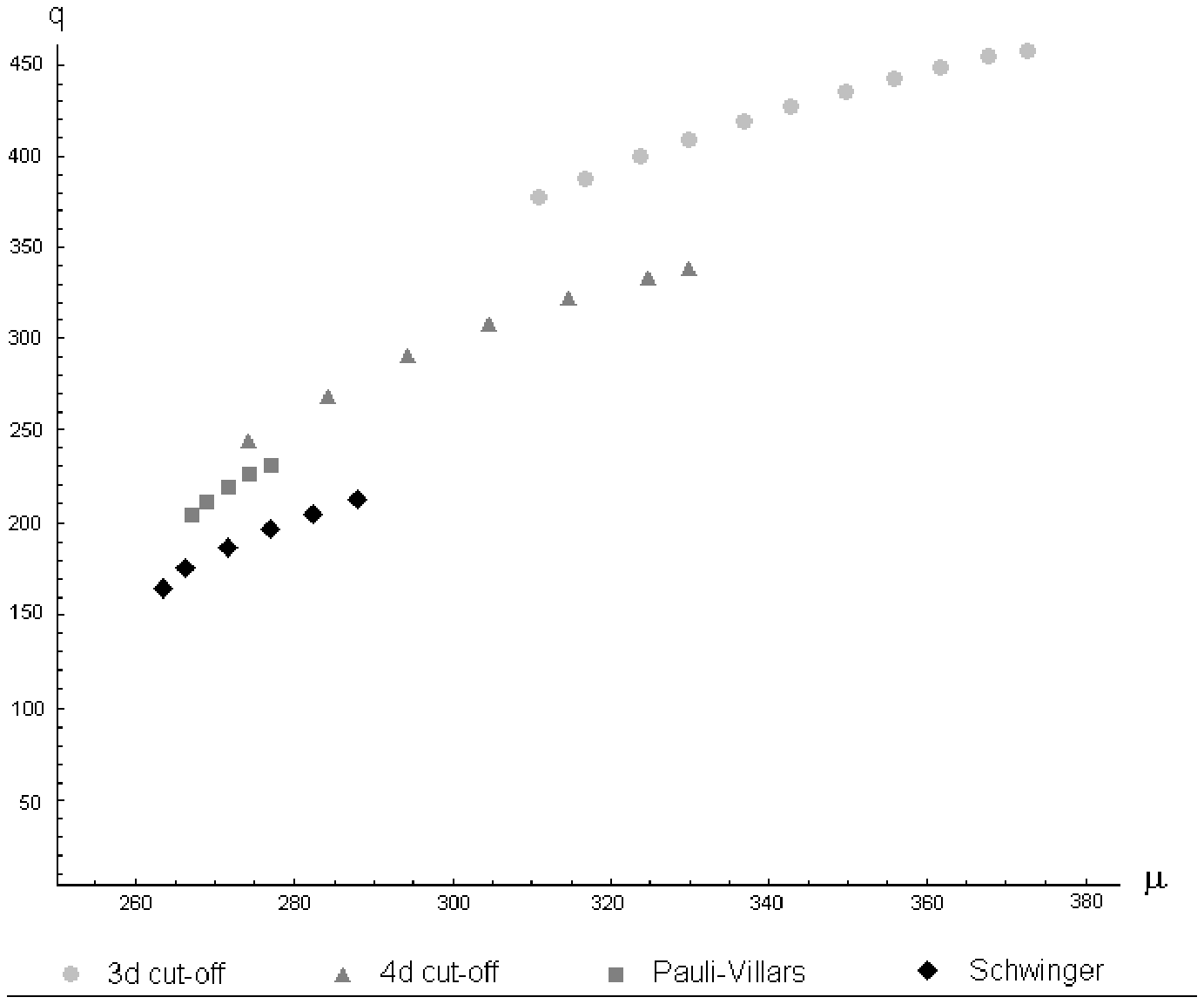}
\end{center}
\caption{The value of the wave vector $q$ as a function of the
chemical potential $\mu$ in the different regularization schemes in MeV
units. The values of $\mu$ for non-zero $q$ describe the range of
the existence of the non-uniform chiral phase.} \label{number}
\end{figure}

Some quantitative features change with the chosen scheme. One can find that schemes cluster in two groups which one can call
"relativistic" schemes (4D, S, PV) and 3D cut-off. However, let us note
that the distinction between relativistic and non-relativistic schemes
has no deep meaning because the thermodynamic system singles out one reference frame.
Comparing the quantitative results, we consider values of constituent mass, wave vector, gap parameter, the strengths of the
first order phase transitions and critical chemical potential. The position of the transition from uniform chiral to nonuniform
chiral phase is the most resistant
against the choice of the regularization scheme. First critical potential changes within the range of 5 per cent for
relativistic schemes, while in the case of 3D within 18 per cent. The position of the second transition, for relativistic
schemes changes within the range of 19 per cent, including 3D cut-off within 34 per cent. For 4D, S, PV schemes the range of
variability of the parameter M value is about 20 per cent at $\mu$ equal to zero, and about 60 per cent at chemical potentials
corresponding with first transition. The range of variability of parameter q value is 50 per cent at chemical potentials
corresponding with first transition. Comparing M and q at second
transition is unsuitable because of changeability of second critical potential. The dependence of the parameters
M and q on the chemical potential is the same
for the different regularizations. With increasing chemical potential, the value of q is growing, the M value is declining.
Relatively the least correlated is the dependence of the parameter $\Delta$ on the regularization scheme. However, again the
value of the gap $\Delta$ is increasing with $\mu$, independent of the regularization choice.
The values of the critical chemical potentials of the phase transitions are given in Table II.
\begin{center}
\begin{table}[h]
\begin{tabular}{ccccc}
    $\mu_c$  & 3D & 4D & S & PV \\
\hline\hline
Ch/NCH    & 0.311 & 0.274 & 0.263 & 0.268 \\
NCH/2SC   & 0.373 & 0.330 & 0.296  &0.281 \\
\hline\hline
\end{tabular}
\caption{Numerical values of the critical chemical potentials $\mu_c$ in MeV
for uniform and non-uniform chiral phase transition (Ch/NCH) and non-uniform chiral
and superconducting phase transition (NCH/2SC) for different regularization
schemes. In 3D, 4D, S schemes the 2SC phase appears already at Ch/NCh phase
transition and both phases coexist. However, the value of the gap parameter is
relatively low in the coexistence region. In the PV scheme, 2SC phase appears only at
the NCh/2SC phase transition.}
\end{table}
\end{center}

\begin{figure}[b]
\begin{center}
\includegraphics[width=.8\textwidth]{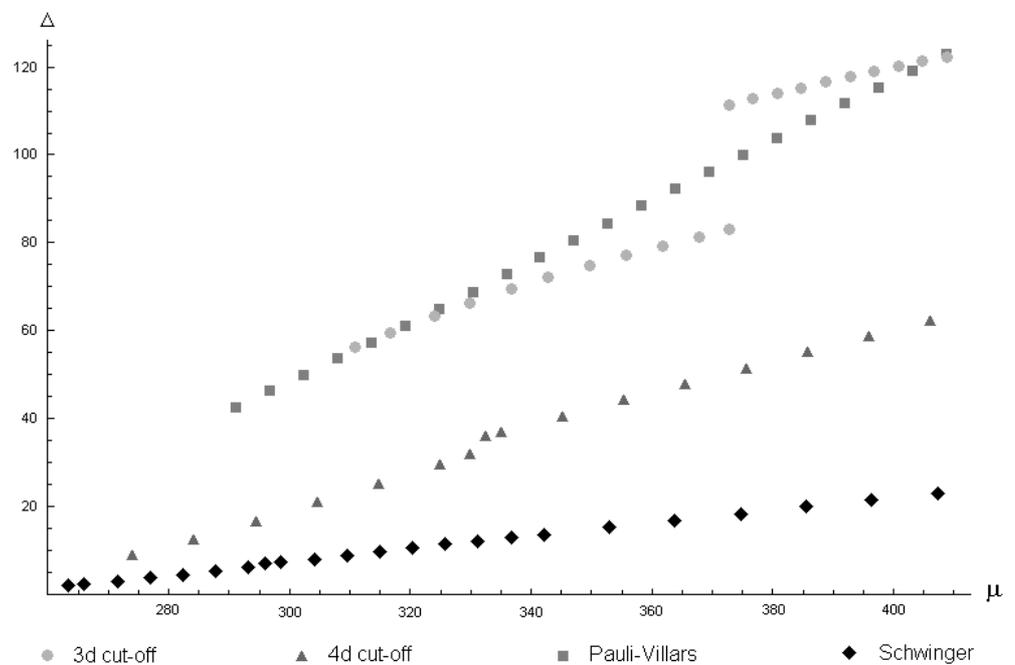}
\end{center}
\caption{The gap parameter $\Delta$ as a function of chemical potential $\mu$
in different regularization schemes in MeV units.} \label{number}
\end{figure}

Strengths of phase transitions depend on the regularization scheme.
The strongest phase transitions are in 3D cut-off, the weakest in the Schwinger
proper time regularization. In the case of the transition to the 2SC phase the jump
of the gap ranges from 28 MeV for 3D cut-off to 1 MeV for Schwinger regularization.
There is still a possibility of the coexistence between chiral and superconducting phase.
It occurs in all schemes with the exception of Pauli - Villars. Any conclusion
which follows from this phenomenon is thus model dependent.
Finally, let us note that the phase diagram depends also on the value of $G^\prime$.
Its influence is the same for all regularization schemes. The larger value of $G^\prime$,
the shorter the range of non-uniform chiral phase.

This behavior is understandable because larger $G^\prime$ strengthens
diquark interaction which dominates over quark - antiquark
interaction. Only in the Pauli - Villars scheme, the non-uniform chiral phase
vanishes for $G^\prime = 0.83 G$, and the phase transition to superconducting
phase at $\mu = 0.26$ GeV takes place directly from the uniform chiral phase.However,
this value of the critical chemical potential is rather low
which questions the physical sensibility to set $G^\prime = 0.83 G$ in PV scheme.

\section{Conclusions}

We perform the analysis of the phase diagram of strongly interacting matter in
the Nambu - Jona-Lasinio model which includes non-uniform chiral phase and superconducting
2SC phase in different regularization schemes. We confirm that the qualitative features
of the phase diagram is independent of considered regularization schemes. The
generic phase diagram is shown in Fig. 4 (in 4D cut-off regularization).

The quantitative results (values of constituent mass, wave vector, critical chemical potential) match satisfactorily
within "relativistic" schemes.
Only results obtained with 3D cut-off differ widely from the others. Unfortunately, there is no general argument
which scheme better suits the task of the phase diagram analysis. Neither the relativistic
approach (thermodynamic systems single out prefered reference frame) nor gauge independence
(the NJL model is not a gauge theory) favor any scheme in the present considerations.
Additionally, the gap parameter $\Delta$ as well as the magnitude of the jump in the gap parameter
shows the clear dependence on the choice of the regularization scheme. The differences appear not only
between the "relativistic" and 3D cut-off schemes but also within the set of "relativistic" regularizations.
These findings tell us that one can set the magnitude of the gap parameter in the MeV scale but its precise value
in the large extand is an unknown quantity. From the other hand the general qualitative pattern that the gap parameter increases with
increasing value of the chemical potential $\mu$ is independent on the choice of the regularization scheme.

\begin{figure}[t]
\begin{center}
\includegraphics[width=.8\textwidth]{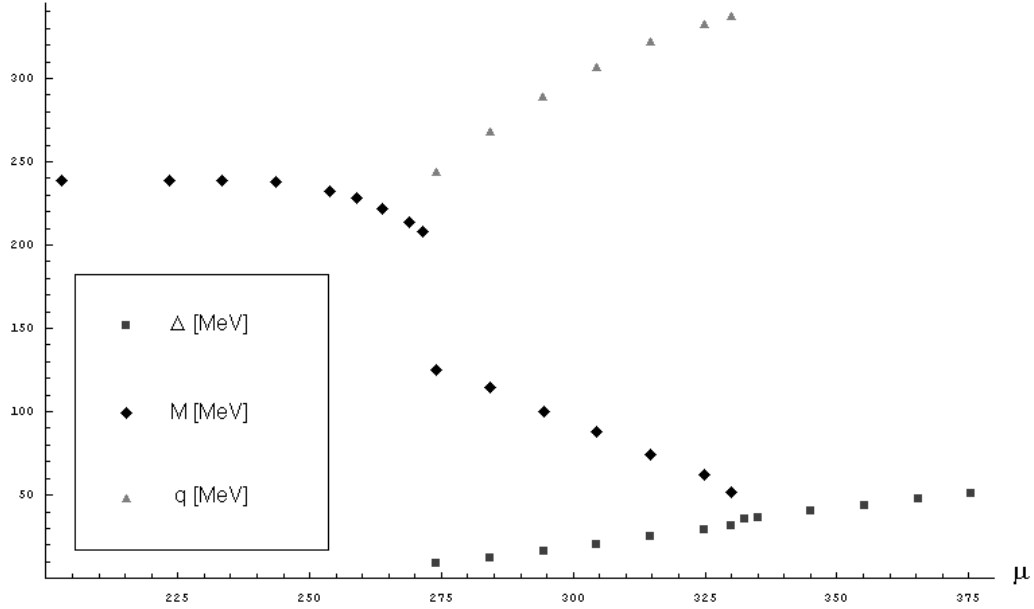}
\end{center}
\caption{The phase diagram at one plot in 4D cut-off regularization schemes (in MeV units).} \label{number}
\end{figure}

The size of the non-uniform phase depends on the relative strength of $G^\prime$ and $G$
coupling constants. The larger $G^\prime$ constant the shorter range of the non-uniform phase.
This conclusion is also independent of the regularization scheme.

Finally, we find that in the Pauli - Villars scheme, in contrast to the other schemes,
there is no coexistence region of the non-uniform and 2SC phases. Thus such
a coexistence remains an open question.
Let us stress at the end that our analysis does not prove that the non-uniform chiral phase
exists. However, it shows that the main features of the phase diagram
which includes non-uniform phase are robust against the choice of the
regularization schemes.

\section{Appendix A}

Two parameters of the NJL model ($G, \Lambda$) are fixed by two physical quantities:
the pion decay constant $f_{\pi}=93$ MeV, and the quark condensate density
$\left\langle \bar{u}u\right\rangle=\left\langle \bar{d}d\right\rangle=$-(250 MeV)$^3$.
These quantities are functions of $M_{0}$ and $ \Lambda$, and
can be calculated in the framework of the NJL model, as was done in \cite{klev}.
Alternatively one can use the decay constant $g_{\rho}$ for the $\rho\rightarrow 2\pi$ process instead of
the quark condensate value, proposed in \cite{blasz}. Now using the self-consistency
condition, $M_{0}$ = $-4G\left\langle \bar{u}u\right\rangle$, that links $G$ and $\Lambda$ with $M_{0}$,
we get values of $G$ and $\Lambda$.

In different regularization schemes, one has
\begin{itemize}
\item \textbf{3D cut-off}
$$
\left\langle\bar{u}u\right\rangle=-6M_{0}\int^{\Lambda}\frac{d^{3}k}{(2\pi)^{3}}{\frac{1}{E_{0}}},\;\;\;
f^{2}_{\pi}=3M_{0}^{2}\int^{\Lambda} \frac{d^{3}k}{(2\pi)^{3}}{\frac{1}{E_{0}^{3}}} .
$$
\item \textbf{4D cut-off}
$$
\left\langle \bar{u}u\right\rangle=-12M_{0}\int^{\Lambda}\frac{d^{4}k_E}{(2\pi)^{4}}{\frac{1}{E_{0}^{2}+k_{0}^{2}}},\;\;\;
f^{2}_{\pi}=12M_{0}^{2}\int^{\Lambda}\frac{d^{4}k_E}{(2\pi)^{4}}{\frac{1}{(E_{0}^{2}+k_{0}^{2})^{2}}} .
$$
\item \textbf{Schwinger}
$$
\left\langle \bar{u}u\right\rangle=-\frac{3}{4\pi^{2}}M_{0}\int^{\infty}_{1/\Lambda^{2}}\frac{d\tau}{\tau}{\exp[-\tau M_{0}^{2}]},\;\;\;
f^{2}_{\pi}=\frac{3}{4\pi^{2}}M_{0}^{2}\int^{\infty}_{1/\Lambda^{2}}\frac{d\tau}{\tau}{exp[-\tau M_{0}^{2}]} .
$$
\item \textbf{Pauli-Villars}
$$
\left\langle\bar{u}u\right\rangle=
\frac{3}{4\pi^{2}}M_{0}\left\{\sum^{3}_{\alpha=1}C_{\alpha}M_{\alpha}^{2}\ln\left[\frac{M^{2}}{M_{\alpha}^{2}}\right]\right\},\;\;\;
f^{2}_{\pi}=\frac{3}{4\pi^{2}}M_{0}^{2}\left\{\sum^{3}_{\alpha=1}C_{\alpha}\ln\left[\frac{M^{2}}{M_{\alpha}^{2}}\right]\right\}
$$
with three conditions impose on regularization parameters
\be
\label{PV_cond}
1+\sum^{3}_{\alpha=1}C_{\alpha} = 0,\;\;\;M^2+\sum^{3}_{\alpha=1}C_{\alpha}M_{\alpha}^{2} = 0,\;\;\;
M^4+\sum^{3}_{\alpha=1}C_{\alpha}M_{\alpha}^{4} = 0.
\ee
Conditions (\ref{PV_cond}) are solved by the formulae:
\be
\label{PV_param}
C_{1}=-3,\ C_{2}=3,\ C_{3}=-1,\ M_{1}^{2}=M^{2}+\Lambda^{2},\ M_{2}^{2}=M^{2}+2\Lambda^{2},\ M_{3}^{2}=M^{2}+3\Lambda^{2}
\ee
\end{itemize}
The values of $G, \Lambda$ for each regularization scheme are given in Table I.

{\bf Acknowledgement}
This research was supported by the MEiN grant no. 1P03B 045 29 (2005-2008).

\end{document}